\documentclass[11pt,twoside]{article}
\usepackage{asp2010}

\resetcounters

\begin{document}

\title{A new HDF5 based raw data model for CCAT}
\author{Reinhold~Schaaf$^1$, Adam~Brazier$^2$, Tim~Jenness$^2$,  Thomas~Nikola$^2$,
and Martin~Shepherd$^3$
\affil{$^1$Argelander-Institut f\"ur Astronomie, Universit\"at Bonn, Auf dem H\"ugel 71, 53121 Bonn, Germany}
\affil{$^2$Center for Radiophysics and Space Research, Cornell University, Ithaca, NY 14853, USA}
\affil{$^3$California Institute of Technology, 1200 E California Blvd, Pasadena, CA 91125, USA}
}

\begin{abstract}
CCAT will be a large sub-millimeter telescope to be built near the
ALMA site in northern Chile. The telescope must support a varied set
of instrumentation including large format KID cameras, a large heterodyne
array and a KID-based direct detection multi-object spectrometer.
We are developing a new raw data model based on HDF5 that can cope with
the expected data rates of order Gbit/s and is flexible enough to hold
raw data from all planned instruments.
\end{abstract}

\section{Introduction}

CCAT
\citep{2012SPIE.8444E..2MW,2013AAS...22115006G}
will be a large sub-millimeter telescope to be built near the ALMA site in northern Chile.
Operating at this altitude results in excellent transparency across
all observing bands from 350\,\micron\ to 2\,mm, including the potential
to observe at 200\,\micron\ in the best weather conditions.

The telescope must support a varied set of instrumentation including
large format KID cameras \citep[SWCam;][]{2014SPIE9153-21},
a large heterodyne array \citep[CHAI;][]{GoldsmithCHAI2012}
and a KID-based direct detection multi-object spectrometer \citep[X-Spec;][]{2014SPIE9153-70}.
CCAT's raw data model must be flexible enough to be used with any of these
instruments.

Since the expected data rates for the instruments are of order Gbit/s,
it was decided to use a
loosely-coupled distributed data acquisition system where instruments
and the telecope control system (TCS)
write out time-series independently of all other systems
but where synchronization is managed by accurate recording of time
stamps. It is up to the data reduction software to take the time
stamps and determine which sequences are related. Each system writes
out time-series using a shared data model and a central data capturer
task \citep{2014SPIE9152-109} collates the individual components and creates a linker file
referencing them. The data capturer does not require highly
synchronized coordination of data writing between systems.

\begin{figure}[t]
\footnotesize
\begin{quote}
\begin{verbatim}
<ObsID>.h5              Group                 Basic metadata
|--OCS                  Group                 observation metadata
|--...                                        other static metadata
|--TCS                  Group (ext. link)     static TCS metadata
|  |--TCS_00001         Group (ext. link)     TCS data (30s)
|   --TCS_00002         Group (ext. link)     TCS data (30s)
|--SWCam350             Group (ext. link)     static SWCam350 metadata
|  |--SWCam350_00001    Group (ext. link)     SWCam350 data (30s)
|   --SWCam350_00002    Group (ext. link)     SWCam350 data (30s)
 --SWCam450             Group (ext. link)     static SWCam450 metadata
   |--SWCam450_00001    Group (ext. link)     SWCam450 data (30s)
    --SWCam450_00002    Group (ext. link)     SWCam450 data (30s)
\end{verbatim}
\end{quote}
\caption{The HDF5 hierarchy of an observation with the 350\,\micron\ and
450\,\micron\ sections of SWCam. The HDF5 root group and other groups
directly below it in the hierarchy (like the shown OCS group) contain
observation-related static metadata. TCS and instrument data and metadata
are stored in separate files, external links are used to establish a
single HDF5 hierarchy for all data in the dataset.}
\label{fig:HdfHierarchy}
\end{figure}

\section{Evaluation of existing raw data models}

We have evaluated the existing data models
MBFITS \citep{2006A&A...454L..25M}, NDF \citep{jennessNDF,P91_adassxxiii} and
LOFAR Data Types \citep{2012ASPC..461..283A}.
In view of CCAT's demands, each of these data models has its
advantages and disadvantages:
MBFITS keeps data from different systems in different files, in accordance with the envisaged
distributed data acquisition scheme; however, although MBFITS was designed and is being used for
continuum and spectral line arrays conceptually similar to CCAT's first-light instruments, the
data model is not flexible enough for new types of instruments like X-Spec.
In contrast, NDF gives data authors wide freedom to design specialized data models on top of the
general NDF model (for example those used for the SCUBA-2
\citep{2013MNRAS.430.2513H} and ACSIS \citep{2009MNRAS.399.1026B} instruments) but lacks
facilities for transparently linking structures across different files or for implementing tabular
time-series data in an efficient manner\footnote{Neither of these issues are fundamental issues
with NDF and could easily be solved by implementing NDF on top of HDF5.}.
The LOFAR Data Types (implemented in HDF5) are a family of related hierarchical data models for raw data and
data products for various LOFAR observing modes; they share common structures for common data and metadata
and allow specialized structures. However, the data models reflect the specific structure of the LOFAR array
and its observation modes too much to be used directly for a single-dish telescope with radically different
observing modes.

\section{Layout of the raw data model }

HDF5 \citep{2011Folk:OHT:1966895.1966900} is a
fundamentally hierarchical format that  matches our design philosophy,
but also supports critical features such as external links between
files and high performance writing of time-series data using the
packet table interface. For these reasons we have chosen to adopt HDF5
as our low-level data format on which to layer our data model.

During an observation, the data capturer, the TCS, and involved
instruments write their data to HDF5 files independently. The
set of these files forms a dataset that contains all data and
metadata of the involved systems during this observation.
In order to avoid excessive file sizes, bulk data from
the TCS and science instruments will be recorded in sequences of data
files which hold chunks of data for 30\,s each.

In order to form a single HDF5 hierarchy from the HDF5 structures in
the files of a dataset, HDF5's external links are used. The result is
a HDF5 hierarchy with basic observation-related metadata at the root
of the hierarchy, and TCS and instrument specific structures further
down the hierarchy.
Each system will write out structures in a standard way
such that the TCS component of a CHAI observation will be identical to
that of an SWCam observation. Furthermore, following the lead from
NDF, structure layouts will be re-used wherever
possible\footnote{Sometimes even using the NDF naming convention when
  that is appropriate.} when
designing the form of instrument-specific structures and, for example,
the time field in every time-series table will use the same name and
format to encourage code re-use and aid in cross-instrument
understanding. There is, however, no requirement for each instrument to
adopt data models that do not fit well with the
needs of the particular instrument.
This approach provides a good compromise between
a well-constrained model and one with sufficient flexibility to cope
with the specific needs of instruments.
Figures~\ref{fig:HdfHierarchy}, \ref{fig:TcsHierarchy}, and \ref{fig:SwcamHierarchy}
illustrate the proposed HDF5 hierarchy.

\begin{figure}[t]
\footnotesize
\begin{quote}
\begin{verbatim}
TCS_00001             Group      TCS data (30s)
|--TCS_300Hz          Group
|  |--Data            Dataset    TCS data sampled at 300Hz
|   --Quality         Group      related quality data
|--TCS_30Hz           Group
|  |--Data            Dataset    TCS data sampled at 30Hz
|   --Quality         Group      related quality data
 --TCS_1Hz            Group
   |--Data            Dataset    TCS data sampled at 1Hz
    --Quality         Group      related quality data
\end{verbatim}
\end{quote}
\caption{The HDF5 sub-hierarchy of the HDF5 hierarchy in
Fig.~\ref{fig:HdfHierarchy} holding 30\,s of TCS data. Since the TCS
samples data at three different frequencies (300\,Hz, 30\,Hz,
and 1\,Hz), sub-structures for each of these frequencies are needed.
Each of these sub-structure comprises a dataset \texttt{Data} with a
table containing timestamps and TCS data, and an additional group
\texttt{Quality} (similar to NDF's \texttt{QUALITY} structure)
allowing storage of a bit mask for each element of the \texttt{Data}
dataset indicating missing or invalid data.}
\label{fig:TcsHierarchy}
\end{figure}

\begin{figure}[t]
\footnotesize
\begin{quote}
\begin{verbatim}
SWCam350                 Group
|--Calibration
|  |--FocalPlane         Group   static metadata
|  |--FlatField          Group   static metadata
|   --BeamModel          Group   static metadata
|--SWCam350_00001        Group   SWCam350 data (30s)
|  |--Housekeeping       Group   biases, pressures, temperatueres...
|  |--Readoutline_001    Group   data from ~1200 detectors
|  |--Readoutline_002    Group   data from ~1200 detectors
|   ...
 --SWCam350_00002        Group   SWCam350 data (30s)
   |...
\end{verbatim}
\end{quote}
\caption{The HDF5 sub-hierarchy of the HDF5 hierarchy in
Fig.~\ref{fig:HdfHierarchy} for data from the 350\,\micron\
section of SWCam. The sub-hierarchy contains structures for static
metadata needed for calibration and for 30\,s chunks of detector data.
The detector data will be time-stamped and transparently compressed.}
\label{fig:SwcamHierarchy}
\end{figure}

HDF5's external links rely on pathnames of the referenced files; since
absolute pathnames are not invariant when files are moved, only relative
pathnames are used. This requires that all files of a
dataset reside in a single directory tree which can be achieved
with mounts and (file-system) symbolic links. It is also likely
that we will adopt the approach of using a distributed file system
such as GPFS.

\pagebreak
\bibliography{P3-1}

\end{document}